%% file: ex_article.tex
\documentclass[review,onefignum,onetabnum]{siamonline171218}

\usepackage{wrapfig}
\usepackage{times}
\input{ex_shared}
\usepackage{subfig}
\usepackage{multirow}
\usepackage{helvet}
\usepackage{courier}
\usepackage{epsf}
\usepackage{epsfig}
\usepackage{mathrsfs}
\usepackage{graphics}
\usepackage{booktabs}
\usepackage{balance}
\usepackage{graphicx}
\usepackage{balance}
\usepackage{epstopdf}
\usepackage{pifont}

\usepackage{amsfonts}

\usepackage{pifont}

\newcommand{\wben}{\beta}
\newcommand{\wdet}{\delta}
\newcommand{\wgain}{w_g}
\newcommand{\techrate}{\tau}
\usepackage{epstopdf}

\DeclareMathSizes{10}{9}{6}{4}
\usepackage{etoolbox}
\ifpdf
\hypersetup{
  pdftitle={TBD},
  pdfauthor={Anonymous}
}
\fi

\begin{document}
\nolinenumbers
\maketitle

\begin{abstract}
Implicit biases occur automatically and unintentionally and are particularly present when we have to make {\em split second} decisions. 
One such situations appears in refereeing, where referees have to make an instantaneous decision on a potential violation. 
In this work we revisit and extend some of the existing work on implicit biases in refereeing. 
In particular, we focus on refereeing in the NBA and examine three different types of implicit bias; (i) home-vs-away bias, (ii) bias towards individual players or teams, and, (iii) racial bias. 
For our study, we use play-by-play data and data from the Last Two Minutes reports the league office releases for games that were within 5 points in the last 2 minutes since the 2015 season. 
Our results indicate that the there is a bias towards the home team - particularly pronounced during the playoffs - but it has been reduced since the COVID-19 pandemic. 
Furthermore, there is robust statistical evidence that specific players benefit from referee decisions more than expected from pure chance. 
However, we find no evidence of negative bias towards individual players, or towards specific teams. 
Finally, our analysis on racial bias indicates the absence of any bias. 
\end{abstract}

\section{Introduction}

Being a referee in sports is without question a very tough job. 
There are decisions that need to be made in literally a split second, and they are able to do it at a very high accuracy. 
On top of this, they have to endure almost constant complaints from the two teams being refereed. 
When having to make decisions this quickly, the human brain has to rely on various heuristics and this is where implicit bias can get into the way of judgement \cite{schirrmeister2020psychological}. 
Referees - like all humans - are not immune to these type of biases and prior work has reported on a variety of similar instances. 
For example, baseball umpires exhibit the gambler's fallacy in the call of pitches, showing a negative auto-correlation in their calls of consecutive ambiguous pitches \cite{chen2016decision}. 
Umpires also exhibit higher error rate when there were 3 balls or 2 strikes (excluding full counts), favoring the call that would not end the at bat \cite{moskowitz2011scorecasting}. 
This is a result of another cognitive shortcut, namely, impact aversion, which is essentially a bias towards doing nothing. 
Price and Wolfers \cite{price2010racial} using foul data from the NBA for the seasons between 1992-2004 found that on average players get called for more fouls when officiated by an opposite-race crew as compared to when being officiated by a same-race crew. 
This study steered a lot of discussion in the league office and in 2010 Pope, Price and Wolfers \cite{pope2018awareness} revisited the question and analyzed data over two 3-year periods, one before the publication of the original study (2003-2006) and one after (2007-2010). 
They found that during the first period there was still a significant racial bias in calling fouls, while this bias was no longer present in the second period. 
This is a valuable finding, since it provides evidence that the knowledge of implicit biases can help in reducing or even eliminating them.  
More recently Mocan and Osborne-Christenson \cite{mocan2022group}, using data from the NBA's last two minute reports (to be described later) did not find any biases with regards to incorrectly called fouls, but there were significant in-group biases with regards to non-called fouls.  

In this work, we revisit and expand some of this literature, examining possible implicit biases in NBA refereeing with respect to: (i) home - vs - away teams, (ii) individual (super star) players or teams, and, (iii) players/referee race. 
We use play-by-play data as well as data from the Last Two Minute (L2M) reports since the 2015 season for our study. 
The L2M reports include a detailed break down of the events that took place during the last 2 minutes of {\em close} games, defined as games within 5 points. 
There is an entry for each call on whether the call was correct or not. 
There is also information about which player/team benefited/disadvantaged from this call. 
Furthermore, there is the same information on missed call. 
For studying the first two types of implicit bias we use the L2M reports and estimate the net whistle gain for a team or player based on the situations where they benefited or were disadvantages. 
We further empirically estimate its statistical significance through Monte Carlo simulations. 
For the racial bias we make use of play-by-play data since the L2M reports do not have information with regards to which referee made the call. 
Even though it is possible to match the L2M data with play-by-play and obtain information about which referee made a call\footnote{Even though, there are several inconsistencies among the two datasets, particularly with regards to the game clock.}, only 3\% of the foul calls (a total of 210) is incorrect. 
On the contrary while there are more incorrect non-calls when it comes to fouls (a total of 1,399 for approximately 12\% of all fouls) it is impossible to know which referee was responsible for the call (e.g., the closest referee). 
Thus, we will use play-by-play data similar to \cite{price2010racial,pope2018awareness}. 
However, we are not relying on foul calls. 
In order to properly analyze these calls we need to consider which player benefited from the call as well, a piece of information missing from prior analysis and one that we do not have for all foul calls from play-by-play data. 
Furthermore, given that only a very small of the fouls called is incorrect, then this could bias the calculations  as most of the calls are correct and therefore, needed to be made. 
Hence, we rely on analyzing the technical fouls called from referees, which are also more subjective compared to foul calls. 

Our main findings can be summarized in the following: 

\begin{itemize}
    \item During the whole period that our data cover there is overall a home-team bias, which is even more pronounced during the playoffs. However, this bias has almost been eliminated since the 2020 season. 
    \item There are specific players that exhibit a statistically significant positive net whistle gain. However, the same is not true for the opposite direction, i.e., players that have a statistically significant negative net whistle gain. 
    \item We do not find evidence of bias in any direction towards individual teams. 
    \item There is not any racial bias observed when analyzing (personal) technical fouls called.
\end{itemize}

The rest of this paper is organized as follows: 
In Section \ref{sec:data-methods} we present the data used and the analysis methods. 
Section \ref{sec:results} presents and discusses the results, while Section \ref{sec:discussion} concludes our work and discusses its limitations and future work.

\section{Data and Methods}
\label{sec:data-methods}

For our study we used the L2M reports data covering the seasons between 2015 (the first season the NBA started releasing the reports) until this past season 2022. 
The data were collected and are made publicly available at the following github repo: \url{https://github.com/atlhawksfanatic/L2M}. 
Each entry in the L2M includes several elements but the ones that we make use of in our analysis are: {\tt committing player}, {\tt disadvantaged player}, {\tt committing side}, {\tt disadvantaged side}, {\tt decision}. 
The decision takes 4 possible values: correct call (CC), incorrect call (IC), incorrect non-call (INC) and correct non-call (CNC). 
While CC, IC and INC decisions are well-defined, CNC decisions are not. 
In theory, every second in the game with no violation is a CNC. 
Hence, the instances included in the reports are subjective and the criteria can change from year-to-year. 
In fact, during the 2015 season there were 6.4 CNC entries per game, while during the 2022 season there were almost 14 CNC entries per game. 
This means that any analysis should not rely on CNC data points since they are not consistent across seasons. 

We also collect the play-by-play data through the NBA API. 
These data provide information for the events that took place during each game, including the technical fouls called. 
We only consider personal technical fouls, that is, we filter out calls like defensive 3 seconds, delay of game etc., that are labeled as technical fouls as well. 
For every technical foul the play-by-play data also provide information for the referee calling it and of course the player receiving it. 
We further collected the demographics of the referees manually, while for players we combined online sources\footnote{\url{https://www.interbasket.net/news/what-percentage-of-nba-players-are-black-how-many-are-white/31018/}} with manual annotation as well. 

{\bf Home Court: }To examine possible home-court biases in refereeing we start by calculating the home team net whistle gain for all the games in the L2M dataset. 
The net whistle gain for the home team consists of two parts, namely, the {\em whistle benefit} and the {\em whistle detriment}. 
The whistle benefit $\wben$ is just the number of INC decisions when the committing side is the home team plus the number of IC decisions when the committing side is the visiting team. 
Similarly, the whistle detriment $\wdet$ for the home team is the number of INC decisions when the committing side is the visiting team plus the number of IC decisions when the committing side is the home team. 
Then the net whistle benefit $\wgain$ for the home team is simply $\wgain = \wben-\wdet$.  
This is essentially the total number of times that the home team benefited from the referee decision. 
If $\wgain > 0$ the home team got overall the ``better whistle'', while if $\wgain < 0$ the visiting team got the better whistle. 

\begin{figure*}[t]

\includegraphics[width=0.33\textwidth]{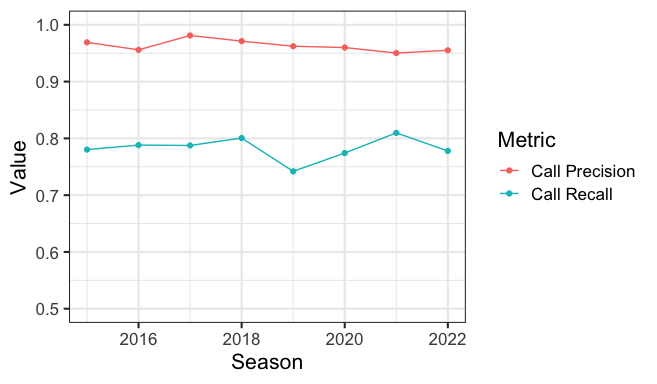}
\includegraphics[width=0.33\textwidth]{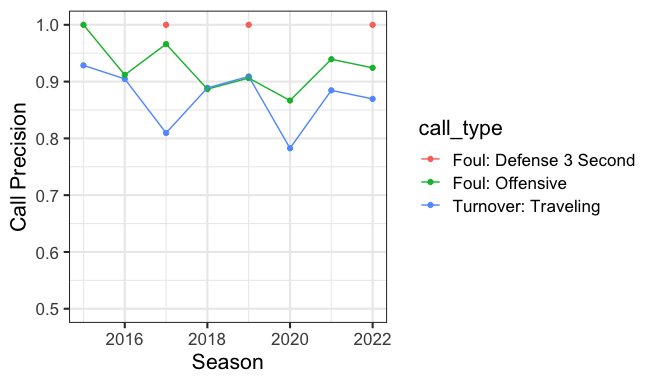}
\includegraphics[width=0.33\textwidth]{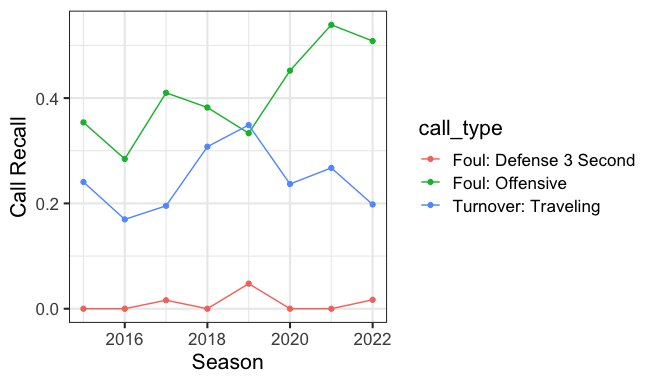}

\caption{Precision and recall overall and of different violation types over the last 8 seasons.}
\label{fig:precision_recall}
\end{figure*}

However, the question is whether $\wgain$ is statistically different than zero or we could have expected this by the stroke of luck. 
In order to answer this question we rely on Monte Carlo simulations. 
In particular, we simulate the decisions on all actual violations and calls in the dataset based on the precision and recall rates of violations. 
We define as the violation calls precision as the ratio: $\dfrac{CC}{CC+IC}$, while the recall of a violation is: $\dfrac{CC}{CC+INC}$. 
Given that not all violation types have the same precision or recall we calculate these metrics separately for each violation type. 
Figure \ref{fig:precision_recall} (left), shows the overall precision and recall for all violations over the seasons covered in our data. 
As we can see when a call is made, this is a true violation with very high probability ($> 95\%$). 
Nevertheless, there is only about 80\% recall rate, that is, about 20\% of the true violations are missed. 
The middle and right parts of Figure \ref{fig:precision_recall} further show the differences in precision and recall rates for different violations (Table 1 provides the precision and recall rates for all types of violations). 
There are some striking observations. 
For instance, almost none of the defensive 3 seconds violations is being called in the last 2 minutes of close contests (low recall), while about 15\% of traveling calls are incorrect (precision $\approx 85\%$). 
Through our discrete event simulations we can estimate the empirical distribution for $\wgain$, $\hat{f}_{\wgain}$, and this will allow us to estimate the empirical p-value for $\wgain$. 
In what follows we provide some details on the core of the simulation engine. 

\begin{figure*} 
\centering
\includegraphics[width=0.6\textwidth]{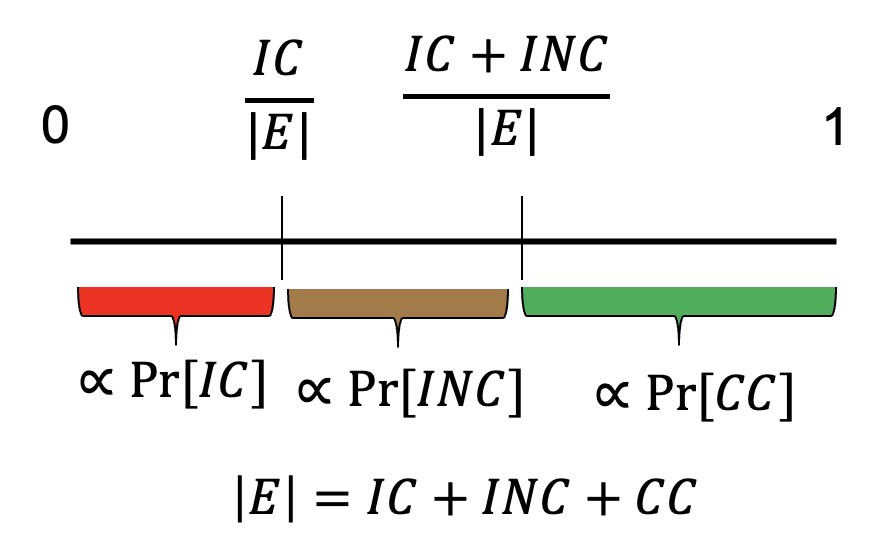}
\caption{Decision boundaries for the simulation of the calls.}
\label{fig:mc_sim}
\end{figure*}

An event for our study is a made call (correct or not) or an actual violation (called or not). 
This means that the total number of events we simulate is essentially $CC+INC+IC$. 
For every event we have to decide whether a correct call was made (CC), or an incorrect call was made (IC) or a violation was erroneously not called (INC). 
The probability of each one of these events is proportional to the corresponding base rate. 
Therefore, for every call we draw a uniformly distributed random number $r$ between 0 and 1 and we have the following decision boundaries (see Figure \ref{fig:mc_sim}): 

\begin{itemize}
    \item If $r \in [0,\dfrac{IC}{IC+INC+CC})$, we have an incorrect call
    \item If $r \in [\dfrac{IC}{IC+INC+CC},\dfrac{IC+INC}{IC+INC+CC})$, we have an incorrect non call
    \item If $r \in [\dfrac{IC+INC}{IC+INC+CC}, 1]$, we have a correct call
\end{itemize}

Given that we should treat each type of violation/call differently, the decision boundaries are different for every type of violation. 
This allows us to control in our simulations for the ``difficulty'' of the violations a team is involved in.

{\bf Player and team-specific: }For examining the presence of player-specific implicit biases by the referees, we use the same method as above. 
However, given that we essentially perform multiple statistical tests - one for each player -- we expect some of them to deem statistically significant results even by chance. 
Therefore, we perform a meta-test to calculate the probability that all of the data points that came out as statistically significant are false positives \cite{pelechrinis2022hot}. 
In particular, under the - realistic in our case - assumption that our tests are not correlated we can use the Binomial distribution for a meta-test. 
With $M$ tests each of which has a probability of $\alpha$ leading to a false positive result, we can estimate the probability of observing at least $r$ positive tests due to chance as: $\sum_{p=r}^{M} \binom{M}{p} \alpha^p (1-\alpha)^{M-p}$. 
If this probability is small, then we can confidently conclude that the non-zero effect sizes observed are not all false positives. 
We follow exactly the same approach for examining team-specific biases. 

{\bf Racial Bias: }The last type of implicit bias that we examine is that of race, which, is an instance of the affect bias/heuristic \cite{slovic2007affect}. 
As aforementioned we will rely on the technical fouls called and in particular we will compare (a) the call rate of technical fouls to players of the same race as the referee $\techrate_{same}$ making the call with, (b) the call rate to players of different race $\techrate_{diff}$. 
This requires the computation of not only the number of technical fouls within and across races, but also of the total minutes that a referee was on the court with players of the same and different race \cite{winston2022mathletics}.  
In order to estimate the statistical significant of the difference $\Delta\techrate=\techrate_{diff}-\techrate_{same}$, we rely again on Monte Carlo simulation. 
In particular, for every referee we estimate their overall call rate per game for technical fouls.  
We then iterate over every game they refereed and perform a two-step simulation. 
First, based on the referee's call rate we simulate the binary decision on whether the referee called a simulated technical foul in the game or not. 
Second, if a technical foul is simulated, the recipient is randomly chosen among the players that took the court in the game. 
The probability of a player receiving the simulated technical foul is proportional to their playing time in the game. 
By repeating this process several times we can obtain the empirical distribution $\hat{f}_{\Delta\techrate}$ under the null hypothesis that there is no racial bias (controlling for the racial composition of players and referees in the various games).  

\footnotesize
\hspace{-0.4in}
\begin{tabular}[t]{lrrr|}
  \hline
  Violation & Precision & Recall & N \\ 
  \hline
Turnover: Traveling & 0.85 & 0.24 & 692 \\ 
 Foul: Personal & 0.97 & 0.90 & 7498 \\ 
Turnover: 8 Second Violation & 0.81 & 0.73 &  35 \\ 
Turnover: Out of Bounds & 0.69 & 0.46 &  47 \\ 
 Foul: Shooting & 0.93 & 0.79 & 4433 \\ 
Stoppage: Out-of-Bounds & 0.91 & 0.97 & 299 \\ 
  Foul: Loose Ball & 0.95 & 0.54 & 1162 \\ 
  Instant Replay: Support Ruling & 0.99 & 1.00 & 858 \\ 
  Foul: Double Personal & 0.50 & 0.38 &  22 \\ 
 Foul: Offensive & 0.91 & 0.40 & 1120 \\ 
Turnover: 24 Second Violation & 0.98 & 0.96 & 347 \\ 
 Instant Replay: Overturn Ruling & 0.98 & 0.99 & 302 \\ 
Foul: Technical & 0.95 & 0.92 & 114 \\ 
   Foul: Double Technical & 0.76 & 0.84 &  24 \\ 
   Ejection: Second Technical & 0.62 & 0.73 &  16 \\ 
 Turnover: Offensive Goaltending & 0.87 & 0.73 &  50 \\ 
  Violation: Kicked Ball & 0.92 & 0.85 & 127 \\ 
 Turnover: 3 Second Violation & 0.65 & 0.10 & 137 \\ 
 Turnover: Backcourt Turnover & 0.84 & 0.81 &  60 \\ 
  Violation: Jump Ball & 0.80 & 0.83 &  29 \\ 
  Violation: Lane & 0.87 & 0.36 &  97 \\ 
 Turnover: 5 Second Inbound & 0.72 & 0.36 &  66 \\ 
 Turnover: Jump Ball Violation & 0.62 & 0.62 &  18 \\ 
 Foul: Flagrant Type 1 & 0.79 & 0.86 &  27 \\ 
 Violation: Defensive Goaltending & 0.94 & 0.81 & 100 \\ 
 Foul: Defense 3 Second & 0.58 & 0.03 & 273 \\ 
Turnover: Lost Ball Possession & 0.55 & 0.60 &  15 \\ 
 Turnover: Double Dribble & 0.74 & 0.44 &  45 \\ 
 Violation: Delay of Game & 0.92 & 0.80 &  79 \\ 
 Turnover: Palming & 0.54 & 0.37 &  25 \\ 
  Turnover: Illegal Screen & 0.55 & 0.50 &  17 \\ 
  Foul: Clear Path & 0.85 & 0.90 &  36 \\ 
  Violation: Double Lane & 0.55 & 0.33 &  23 \\ 
  \hline
  \end{tabular}
  \begin{tabular}[t]{lrrr}
  \hline
 Violation & Precision & Recall & N \\ 
  \hline
Turnover: Stepped out of Bounds & 0.92 & 0.77 & 169 \\ 
Turnover: Kicked Ball Violation & 0.55 & 0.55 &  16 \\ 
Foul: Away from Play & 0.84 & 0.45 & 117 \\ 
Foul: Personal Take & 0.99 & 0.99 & 702 \\ 
Foul: Punching & 0.55 & 0.60 &  15 \\ 
 Stoppage: Other & 0.44 & 0.57 &  12 \\ 
Foul: Delay Technical & 0.71 & 0.79 &  25 \\ 
  Turnover: 10 Second Violation & 0.50 & 0.42 &  17 \\ 
  Turnover: Discontinue Dribble & 0.40 & 0.22 &  24 \\ 
  Violation: Other & 0.43 & 0.55 &  19 \\ 
Instant Replay: Support & 0.55 & 0.67 &  14 \\ 
   Foul: Inbound & 0.50 & 0.42 &  17 \\ 
Turnover: Lane Violation & 0.67 & 0.71 &  19 \\ 
 Turnover: 5 Second Violation & 0.86 & 0.53 &  78 \\ 
   Turnover: Inbound Turnover & 0.55 & 0.38 &  21 \\ 
  Turnover: Punched Ball & 0.44 & 0.50 &  13 \\ 
  Instant Replay: Overturn & 0.50 & 0.62 &  13 \\ 
   Turnover: Illegal Assist & 0.44 & 0.50 &  13 \\ 
 Turnover: Lost Ball Out of Bounds & 0.94 & 0.96 & 202 \\ 
  Violation: Free Throw & 0.50 & 0.62 &  13 \\ 
  Turnover: Out of Bounds - Bad Pass Turn & 0.97 & 0.99 & 208 \\ 
 Foul: Shooting Foul & 0.44 & 0.57 &  12 \\ 
 Foul: Hanging Technical & 0.44 & 0.50 &  13 \\ 
  Foul: Offensive Charge & 0.64 & 0.75 &  17 \\ 
   Foul: Personal Block & 0.44 & 0.50 &  13 \\ 
 Foul: Shooting Block & 0.44 & 0.57 &  12 \\ 
Turnover: Bad Pass & 0.62 & 0.73 &  16 \\ 
 Turnover: Foul & 0.50 & 0.62 &  13 \\ 
Turnover: Lost Ball & 0.64 & 0.75 &  17 \\ 
 Free Throw Technical & 0.44 & 0.50 &  13 \\ 
  Stoppage: TimeOut & 0.54 & 0.64 &  17 \\ 
   Stoppage: Clock & 0.55 & 0.60 &  15 \\ 
Foul: Defensive 3 Second & 0.44 & 0.50 &  13 \\ 
   \hline
\label{tab:violations}
\end{tabular}

Table 1: Precision and recall of different types of violation.

\normalsize

\section{Results}
\label{sec:results}

In this section we are going to present our analysis results. 

\subsection{There is referee home court bias but it is reducing}
\label{sec:hca}

We start by looking at the L2M data and estimating the net whistle gain for the home team during the whole period covered in our data. 
Table 2 presents the results, where we can see that there is overall a statistically significant home court referee bias, with the home team having benefited in approximately 146 situations more than expected. 
This corresponds to an 1.2 percentage units difference between the home and visiting team. 
Furthermore, as we can see the home court bias is much higher during the playoffs. 
Given that home court referee bias is part of the home court advantage (HCA), which has been linked to the home team fans, we wanted to examine separately the seasons during/after the COVID-19 pandemic. 
The NBA finished the 2020 season in a bubble with no fans, and started the 2021 season in empty arenas. 
In fact, most of the teams didn't start having fans at limited capacity until the middle of that season and only reached arenas with fans closer to capacity during the playoffs. 
As we can see from Table 2 the home court referee bias appears to have disappeared during these seasons! 
This is in fact in agreement with the trend of the overall home court advantage as estimated from team regression ratings. 
In particular, based on the Sagarin ratings \cite{sagarin} the home court advantage between 2015-2019 was 2.74 points, while between 2020-2022 it dropped to 1.75 points. 

So overall, we see the presence of a home team referee bias, particularly pronounced during the playoffs. 
However, this has started disappearing after the 2020 seasons. 
It remains to be seen whether this has been an artifact of empty arenas during the COVID-19 pandemic, or whether the trend will continue. 
One other change that took place during the 2020 season, was the introduction of coaches challenge, where a coach can contest one call per game. 
This triggers an automatic review and the call can change. 
When NFL introduced a similar system the win percentage of the home teams dropped from 58.5\% to 56\% \cite{moskowitz2011scorecasting}. 
However, in the NFL coaches can have up to 3 challenges, while in the NBA it is strictly 1. 
Therefore, the impact is expected to be overall smaller, but nevertheless there are several anecdotes supporting its possible impact on the home court advantage. 
For example, during the very first week of this new rule Portland won in Dallas due to a coaches challenge that overturned a foul called five seconds before the end of the game against Portland that most probably would have given the win to the home team\footnote{\url{https://www.dallasnews.com/sports/mavericks/2019/10/29/controversial-overturned-call-in-mavs-blazers-shows-coachs-challenge-is-a-work-in-progress-with-an-obvious-flaw/}}.

\begin{table}[ht]
\centering
\begin{tabular}{rllrrr}
  \hline
  seasons & season type & p-val & $\wgain - \mathbf{E}[\hat{f}_{\wgain}]$ (\%) \\
  \hline
    2015-2022 & regular & 0.03 & 107.7 (1.2\%) \\ 
  2015-2022 & playoffs & $<$0.01 & 47.86 (7\%) \\ 
  2015-2022 & both & $<$0.01 & 145.55 (1.6\%)\\
2015-2019 & regular & 0.02 & 97.45 (1.5\%) \\
  2015-2019 & playoffs & $<$0.01 & 41.12 (9.4\%)\\ 
 2015-2019 & both & $<$0.01 & 142.15 (2.2\%)\\ 
  2020-2022 & regular & 0.45 & 2.97 (0.02\%)\\ 
 2020-2022 & playoffs & 0.26 & 6.26 (2\%) \\ 
2020-2022 & both & 0.49 & 8.94 (0.02\%) \\ 

   \hline
\end{tabular}
\caption{The home court bias has reduced since the 2020 season, which is the season of the COVID-19 pandemic.}
\label{tab:hca}
\end{table}


\subsection{There is player-specific bias, but only positive. There is no team-specific bias }
\label{sec:player_bias}

Next we examine the net whistle gain for individual players over the seasons covered from the L2M data. 
We repeat the same process as for the home court referee bias, but now focusing on individual players. 
We only use in our analysis players that have been involved in at least 100 calls/missed calls over the whole 8-year period (this corresponds to the top 10th percentile). 
This provides us with a total of 106 players. 
Also when estimating the base call/miss rate for a violation type we filter out the data of the specific player we simulate. 
Table 3 shows the results for all players where we can see that there are 12 players that exhibit a statistically significant positive net whistle gain (at the 5\% significance level). 
Using the binomial metatest aforementioned, there is an approximately 7-in-1,000 chance that all of these 12 instances are false positives. 
Therefore, we can say with quiet some confidence that there are specific players that get a ``better whistle'' than expected. 
We can also see that most of these players are all-stars, all-NBA and/or all-defensive NBA players (e.g., Dwyane Wade, Chris Paul, Carmelo Anthony, Karl-Anthony Towns, Jayson Tatum, Andre Drummond, Hassan Whiteside, Patrick Beverley). 
We also looked at the opposite direction, i.e., whether there are players that consistently get a ``worse whistle'' than expected. 
There is a total of 7 players that exhibit a statistically significant negative net whistle gain. 
However, the probability of all of these 7 instances being false positives is non-negligible and equal to 28\%. 

Turning to the team-specific analysis, Table 4 depicts the results, where as we can see there are a few (3) teams that have a positive net whistle gain and a few (3) teams that have a negative net whistle gain. 
However, the probability that all of these cases are false positives, is non-negligible as well (19\%). 
Overall, we can say that the data support the presence of a player-specific referee bias. 
However, it is only in one direction, that is, specific players benefiting more than expected. 
Furthermore, the composition of the group of players that exhibit the positive net whistle gain points to a bias towards ``star'' players. 
Nevertheless, given the fact that there are other star players that do not experience the same net benefit it is hard to argue that this is explicitly the reason behind any implicit bias observed. 
Finally, there were no strong evidence of team-specific bias.

\subsection{There is no evidence of racial bias observed among NBA referees}
\label{sec:racial}
\begin{figure}
    \centering
    \includegraphics[scale=0.45]{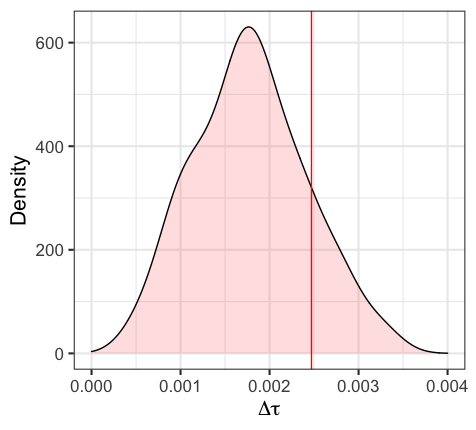}
    \caption{The difference in the personal technical fouls call rate between same and different referee-player race is not statistically different than the one expected by random chance. }
    \label{fig:racial}
    \vspace{-0.2in}
\end{figure}

Lastly we examine the presence of racial bias in refereeing decisions. 
Given that 92\% of both the referees and players in our data are white or African American, we focus on these two racial groups in our analysis. 
We also filter out the games for which we do not have information for all the referees. 
These situations correspond to about 3.6\% of the games. 
In our dataset, there are 5,419 (personal) technical fouls called. 
There were 0.0204 technical fouls per 48 mins called from referees to opposite race players, while, referees called 0.0182 techs per 48 mins to players of the same race. 
So overall, referees called 0.0022 more technical fouls per 48 minutes to players of the opposite race as compared to players the same race as them. 
This difference by itself, even if statistically significant, is hard to be qualified as racial bias, since it corresponds to 1 more technical foul per 450 games approximately. 
Furthermore, we estimated the distribution of tech call rate difference $\Delta\techrate$ through simulating the technical fouls as described earlier. 
Figure \ref{fig:racial} presents the distribution, while the vertical line corresponds to the actual tech call rate difference obtained from the real data. 
In our simulations, we obtained a value of $\Delta\techrate>0.0022$ in 33\% of the cases, indicating that even the small difference observed is not statistically significant. 
This is in agreement with the latest study by Pope, Price, and Wolfers \cite{pope2018awareness}, providing  additional evidence, that is, through examination of different violation calls, for the absence of implicit racial bias by the NBA referees. 

\section{Discussion and Conclusions}
\label{sec:discussion}
In this work we analyzed L2M and play-by-play data from the NBA to analyze a number of different implicit biases that the referees might exhibit. 
We started by looking at the home court bias and we found that while over the past 7 years there is a robust home court advantage, this has been in the decline over the last few years. 
However, given that this period overlaps with the COVID-19 pandemic, and the absence of fans from the arenas, it remains to be seen whether this observation is a trend or an anomaly. 
We then examined the possibility of player and team-specific bias. 
Our analysis indicates that there is evidence for the presence of a bias that is driven by players (not teams) and only in the positive direction (i.e., specific players benefiting more than expected from the calls or non calls). 
Finally, we examined the presence of racial bias in the referee decisions using the personal technical fouls called as our proxy, and we did not find any evidence of racial bias. 
A key part of our analysis is the simulation of the calls/violations recorded in the L2M data. 
This requires the estimation of the call base rates for each violation type. 
These base rates might be noisy when we have very little data points for a given type of violation. 
While this in general can be problematic, in our case we do not expect this to affect our results since these violation types will also not appear in the simulations frequently. 
Nevertheless, an alternative is to use the  Bayesian average for the decision boundaries.  

Some of these findings could provide insights for other areas where similar implicit bias might appear. 
For example, referees of scientific work (e.g., research grants, research papers etc.) might also exhibit biases towards specific scientists and thus, policies/procedures should be put in place to avoid it (e.g., double blind reviews).  
As another example, the home court bias could potentially extend to areas like judicial trials (for which research has already shown other types of implicit biases). 
A judge who is familiar with a defense attorney (i.e., the attorney ``plays'' in home court, literally) may be more willing to listen carefully to the attorney’s arguments and motions. 
Here is were true randomization in judge/courthouse assignment can help, and this is exactly what US courts and courts abroad claim to do. 
However, there is evidence that the assignment is not always fully random. 
For example, Huther and Kleiner  \cite{kleiner2022judges} by analyzing bankruptcy fillings between 2010-2020 found that judge assignment is predicted by the lending decisions of hedge funds. 

\scriptsize
\begin{tabular}[t]{|rl||lr|}
  \hline
 & player &  $\wgain - \mathbf{E}[\hat{f}_{\wgain}]$ (\%)  & pval \\ 
  \hline
1 & Harrison Barnes & 2.88(1.93\%) & 0.32 \\ 
  2 & Isaiah Thomas & -1.295(-1.04\%) & 0.65 \\ 
  3 & Stephen Curry & 4.975(2.29\%) & 0.23 \\ 
  4 & Danilo Gallinari & 2.75(2.43\%) & 0.33 \\ 
  5 & James Harden & 10.065(2.48\%) & 0.14 \\ 
  6 & LeBron James & -9.35(-3.38\%) & 0.92 \\ 
  7 & Brook Lopez & 2.145(1.69\%) & 0.34 \\ 
  8 & Andrew Wiggins & -8.44(-4.72\%) & 0.96 \\ 
  9 & JJ Redick & -0.99(-0.82\%) & 0.58 \\ 
  10 & DeAndre Jordan & 1.46(1.11\%) & 0.39 \\ 
  11 & Paul Millsap & 1.56(1.11\%) & 0.40 \\ 
  12 & Jeff Teague & 2.035(1.65\%) & 0.36 \\ 
   13 & {\bf Dennis Schroder }& 10.965(5.96\%) & 0.02 \\
  14 & LaMarcus Aldridge & -0.43(-0.27\%) & 0.56 \\
  15 & Nicolas Batum & -4.99(-4.94\%) & 0.91 \\ 
  16 & Wesley Matthews & -2.125(-1.7\%) & 0.74 \\ 
  17 & Damian Lillard & -4.165(-1.46\%) & 0.79 \\ 
  18 & {\bf Dwyane Wade} & 9.72(6.89\%) & 0.04 \\ 
  19 & {\bf Hassan Whiteside} & 10.44(9.67\%) & 0.01 \\ 
  20 & Anthony Davis & 2.74(1.36\%) & 0.33 \\ 
  21 & Russell Westbrook & 3.41(0.97\%) & 0.33 \\ 
  22 & Kyrie Irving & 2.67(1.47\%) & 0.36 \\ 
  23 & Marcus Morris & 6.57(4.73\%) & 0.10 \\ 
  24 & PJ Tucker & 2.905(1.73\%) & 0.29 \\ 
  25 & Eric Bledsoe & 3.995(3.1\%) & 0.23 \\ 
  26 & Mike Conley & -5.405(-3.02\%) & 0.89 \\ 
  27 & Marc Gasol & 5.8(3.67\%) & 0.17 \\ 
  28 & DeMarcus Cousins & -1.59(-1.02\%) & 0.66 \\ 
  29 & Tobias Harris & 4.035(2.48\%) & 0.21 \\ 
  30 & Bradley Beal & 1.65(0.72\%) & 0.44 \\ 
  31 & John Wall & -2.27(-1.47\%) & 0.67 \\ 
  32 & George Hill & -0.13(-0.1\%) & 0.55 \\ 
  33 & Kent Bazemore & 0.58(0.44\%) & 0.50 \\ 
  34 & Marcus Smart & 7.615(3.95\%) & 0.09 \\ 
  35 & Khris Middleton & -5.84(-2.86\%) & 0.84 \\ 
  36 & Jerami Grant & -1.51(-1.14\%) & 0.68 \\ 
  37 & Robert Covington & 0.38(0.26\%) & 0.56 \\ 
  38 & Kawhi Leonard & 0.70(0.45\%) & 0.47 \\ 
  39 & Elfrid Payton & -7.94(-7.86\%) & 0.97 \\ 
  40 & Gordon Hayward & -7.495(-5.77\%) & 0.95 \\ 
  41 & Rudy Gobert & 1.215(0.49\%) & 0.47 \\ 
  42 & Will Barton & -7.605(-5.21\%) & 0.95 \\ 
  43 & Zach LaVine & 0.65(0.29\%) & 0.55 \\ 
  44 & Giannis Antetokounmpo & -3.175(-1.04\%) & 0.68 \\ 
  45 & Reggie Jackson & 1.26(0.64\%) & 0.41 \\ 
  46 & Jae Crowder & 0.43(0.34\%) & 0.53 \\ 
  47 & Kentavious Caldwell-Pope & 2.63(2.05\%) & 0.30 \\ 
  48 & {\bf Andre Drummond} & 12.745(7.92\%) & 0.01 \\ 
  49 & Kemba Walker & -2.425(-0.99\%) & 0.68 \\
  50 & CJ McCollum & -1.63(-0.95\%) & 0.62 \\ 
  51 & DeMar DeRozan & -3.35(-1.08\%) & 0.69 \\ 
  52 & Victor Oladipo & -0.29(-0.21\%) & 0.56 \\ 
  53 & Jimmy Butler & 3.545(1.33\%) & 0.33 \\ 
  \hline
  \end{tabular}
  \begin{tabular}[t]{rl||lr|}
  \hline
 & player &  $\wgain - \mathbf{E}[\hat{f}_{\wgain}]$ (\%)  & pval \\ 
  \hline
  54 & Dwight Howard & 7.05(6.91\%) & 0.09 \\ 
  55 & Nikola Vucevic & -4.525(-2.85\%) & 0.83 \\ 
  56 & Kyle Lowry & 8.48(3.64\%) & 0.08 \\ 
  57 & Tim Hardaway Jr. & 3.225(3.1\%) & 0.23 \\ 
  58 &{\bf Chris Paul} & 10.905(3.97\%) & 0.04 \\ 
  59 & Blake Griffin & 1.35(0.83\%) & 0.42 \\ 
  60 & Jrue Holiday & -6.145(-3.09\%) & 0.91 \\ 
  61 & Draymond Green & 5.585(3.12\%) & 0.14 \\ 
  62 & Al Horford & 6.06(3.94\%) & 0.15 \\ 
  63 & Kevin Durant & -3.225(-1.84\%) & 0.77 \\ 
  64 & Paul George & 2.25(1\%) & 0.40 \\ 
  65 & Jonas Valanciunas & 0.835(0.75\%) & 0.47 \\ 
  66 & Aaron Gordon & -2.035(-1.88\%) & 0.70 \\ 
  67 & {\bf Steven Adams} & 10.65(6.16\%) & 0.04 \\ 
  68 & {\bf Carmelo Anthony} & 6.825(6.5\%) & 0.05 \\ 
  69 & Evan Fournier & -4.145(-2.84\%) & 0.84 \\ 
  70 & Ricky Rubio & -0.96(-0.62\%) & 0.64 \\ 
  71 & Julius Randle & -5.48(-2.19\%) & 0.82 \\ 
  72 & Kristaps Porzingis & -1.45(-1.25\%) & 0.66 \\ 
  73 & {\bf Karl-Anthony Towns} & 13.295(6.04\%) & 0.05 \\ 
  74 & {\bf Cody Zeller} & 9.805(9.43\%) & 0.02 \\ 
  75 & Goran Dragic & 1.88(1.17\%) & 0.39 \\ 
  76 & {\bf Mason Plumlee} & 11(10.28\%) & 0.00 \\ 
  77 & Serge Ibaka & 5.855(5.32\%) & 0.12 \\ 
  78 & {\bf Patrick Beverley} & 9.66(8.7\%) & 0.01 \\ 
  79 & Nikola Jokic & -12.485(-4.59\%) & 0.95 \\ 
  80 & Gary Harris & 2.715(2.45\%) & 0.30 \\ 
  81 & Jusuf Nurkic & 4.775(4.01\%) & 0.14 \\ 
  82 & Devin Booker & 0.79(0.33\%) & 0.47 \\ 
  83 & Bojan Bogdanovic & -1.785(-1.65\%) & 0.69 \\ 
  84 & Myles Turner & 1.06(0.88\%) & 0.47 \\ 
  85 & D'Angelo Russell & 0.62(0.55\%) & 0.50 \\ 
  86 & Spencer Dinwiddie & -1.67(-1.11\%) & 0.68 \\ 
  87 & Josh Richardson & -2.025(-1.35\%) & 0.70 \\ 
  88 & Joel Embiid & -2.38(-1.03\%) & 0.67 \\ 
  89 & Brandon Ingram & -2.235(-1.73\%) & 0.75 \\ 
  90 & Jamal Murray & -8.385(-6.35\%) & 0.97 \\ 
  91 & Kelly Oubre & 4.73(4.68\%) & 0.17 \\ 
  92 & Malcolm Brogdon & 0.875(0.84\%) & 0.48 \\ 
  93 & Buddy Hield & -1.04(-0.61\%) & 0.63 \\ 
  94 & Caris LeVert & -7.33(-6.85\%) & 0.98 \\ 
  95 & Jaylen Brown & -0.01(-0.01\%) & 0.56 \\ 
  96 & Fred VanVleet & 2.645(2.62\%) & 0.31 \\ 
  97 & Ben Simmons & -0.585(-0.48\%) & 0.63 \\ 
  98 & De'Aaron Fox & 2.895(1.84\%) & 0.36 \\ 
  99 & {\bf Jayson Tatum} & 8.365(4.78\%) & 0.04 \\ 
  100 & Donovan Mitchell & -4.815(-2.75\%) & 0.86 \\ 
  101 & Bam Adebayo & 7.42(5.38\%) & 0.08 \\ 
  102 & Domantas Sabonis & 6.075(5.15\%) & 0.14 \\ 
  103 & Pascal Siakam & -5.895(-3.88\%) & 0.89 \\ 
  104 & Luka Doncic & -6.18(-5.72\%) & 0.94 \\ 
  105 & Trae Young & 2.16(1.61\%) & 0.42 \\ 
  106 & Ja Morant & 2.15(1.81\%) & 0.37 \\ 
   \hline
\end{tabular}
\label{tab:players}
\normalsize
\hspace{0.2in}

Table 3: Net Whistle Gain for individual players.

\hspace{0.2in}

\begin{tabular}{|rllr|}
  \hline
 & Team & $\wgain - \mathbf{E}[\hat{f}_{\wgain}]$ (\%) & pval \\ 
  \hline
1 & GSW & 3.73(0\%) & 0.41 \\ 
  2 & BOS & 17.79(0.01\%) & 0.14 \\ 
  3 & NOP & 5.26(0\%) & 0.31 \\ 
  4 & DEN & -26.46(-0.02\%) & 0.97 \\ 
  5 & HOU & 6(0.01\%) & 0.32 \\ 
  6 & CLE & -17.36(-0.02\%) & 0.92 \\ 
  7 & BKN & -19.3(-0.02\%) & 0.92 \\ 
  8 & MIN & 13.89(0.01\%) & 0.14 \\ 
  9 & LAC & 10.18(0.01\%) & 0.30 \\ 
  10 & ATL & -6.37(-0.01\%) & 0.67 \\ 
  11 & CHI & -24.7(-0.02\%) & 0.94 \\ 
  12 & WAS & 13.27(0.01\%) & 0.15 \\ 
  13 & CHA & -0.93(0\%) & 0.48 \\ 
  14 & LAL & -7.17(-0.01\%) & 0.76 \\ 
  15 & MEM & 31.25(0.03\%) & 0.00 \\ 
  \hline
  \end{tabular}
  \begin{tabular}{rllr|}
  \hline
 & Team & $\wgain - \mathbf{E}[\hat{f}_{\wgain}]$ (\%) & pval \\ 
  \hline
  16 & POR & 17.75(0.02\%) & 0.14 \\ 
  17 & MIA & 25.24(0.02\%) & 0.04 \\ 
  18 & PHI & -9.79(-0.01\%) & 0.75 \\ 
  19 & OKC & 22.06(0.02\%) & 0.07 \\ 
  20 & PHX & -0.37(0\%) & 0.48 \\ 
  21 & SAC & 7.99(0.01\%) & 0.28 \\ 
  22 & ORL & -9.34(-0.01\%) & 0.79 \\ 
  23 & MIL & -29.02(-0.03\%) & 1.00 \\ 
  24 & IND & 22.16(0.02\%) & 0.06 \\ 
  25 & NYK & -19.25(-0.02\%) & 0.92 \\ 
  26 & DET & 24.56(0.02\%) & 0.04 \\ 
  27 & UTA & -30.14(-0.03\%) & 1.00 \\ 
  28 & SAS & 16.26(0.02\%) & 0.14 \\ 
  29 & DAL & 6.02(0.01\%) & 0.29 \\ 
  30 & TOR & 16.27(0.01\%) & 0.15 \\ 
   \hline
\end{tabular}

\hspace{0.2in}

Table 4: Net Whistle Gain for different teams.

\bibliographystyle{siamplain}
\bibliography{references}

\end{document}

%% file: ex_shared.tex
\usepackage{lipsum}
\usepackage{amsfonts}
\usepackage{graphicx}
\usepackage{epstopdf}
\usepackage{algorithmic}
\ifpdf
  \DeclareGraphicsExtensions{.eps,.pdf,.png,.jpg}
\else
  \DeclareGraphicsExtensions{.eps}
\fi

\usepackage{enumitem}
\setlist[enumerate]{leftmargin=.5in}
\setlist[itemize]{leftmargin=.5in}

\newsiamremark{remark}{Remark}
\newsiamremark{hypothesis}{Hypothesis}
\crefname{hypothesis}{Hypothesis}{Hypotheses}
\newsiamthm{claim}{Claim}

\title{Implicit Biases in Refereeing: Lessons from NBA Referees}

\author{Konstantinos Pelechrinis\thanks{School of Computing and Information, University of Pittsburgh 
 }}

\usepackage{amsopn}

\makeatletter
\newcommand*{\addFileDependency}[1]{
  \typeout{(#1)}
  \@addtofilelist{#1}
  \IfFileExists{#1}{}{\typeout{No file #1.}}
}
\makeatother



%% file: ex_article.bbl
\begin{thebibliography}{10}

\bibitem{chen2016decision}
{\sc D.~L. Chen, T.~J. Moskowitz, and K.~Shue}, {\em Decision making under the
  gambler’s fallacy: Evidence from asylum judges, loan officers, and baseball
  umpires}, The Quarterly Journal of Economics, 131 (2016), pp.~1181--1242.

\bibitem{kleiner2022judges}
{\sc N.~Huther and K.~Kleiner}, {\em Are judges randomly assigned to chapter 11
  bankruptcies? not according to hedge funds}, SSRN (January 27, 2022),
  (2022).

\bibitem{mocan2022group}
{\sc N.~H. Mocan and E.~Osborne-Christenson}, {\em In-group favoritism and peer
  effects in wrongful acquittals: Nba referees as judges}, tech. report,
  National Bureau of Economic Research, 2022.

\bibitem{moskowitz2011scorecasting}
{\sc T.~Moskowitz and L.~J. Wertheim}, {\em Scorecasting: The hidden influences
  behind how sports are played and games are won}, Crown Archetype, 2011.

\bibitem{pelechrinis2022hot}
{\sc K.~Pelechrinis and W.~Winston}, {\em The hot hand in the wild}, PloS one,
  17 (2022), p.~e0261890.

\bibitem{pope2018awareness}
{\sc D.~G. Pope, J.~Price, and J.~Wolfers}, {\em Awareness reduces racial
  bias}, Management Science, 64 (2018), pp.~4988--4995.

\bibitem{price2010racial}
{\sc J.~Price and J.~Wolfers}, {\em Racial discrimination among nba referees},
  The Quarterly journal of economics, 125 (2010), pp.~1859--1887.

\bibitem{sagarin}
{\em Sagarin team ratings}, 2022, \url{http://sagarin.com}.

\bibitem{schirrmeister2020psychological}
{\sc E.~Schirrmeister, A.-L. G{\"o}hring, and P.~Warnke}, {\em Psychological
  biases and heuristics in the context of foresight and scenario processes},
  Futures \& Foresight Science, 2 (2020), p.~e31.

\bibitem{slovic2007affect}
{\sc P.~Slovic, M.~L. Finucane, E.~Peters, and D.~G. MacGregor}, {\em The
  affect heuristic}, European journal of operational research, 177 (2007),
  pp.~1333--1352.

\bibitem{winston2022mathletics}
{\sc W.~L. Winston, S.~Nestler, and K.~Pelechrinis}, {\em Mathletics: How
  gamblers, managers, and fans use mathematics in sports (second edition)},
  Princeton University Press, 2022.

\end{thebibliography}
